\documentstyle[psfig,conf_iap,]{article}
\begin{document}
\heading{%
%
Interstellar Media in the Magellanic Clouds\\ 
and other Local Group Dwarf Galaxies \\
%
} 
\par\medskip\noindent
\author{%
Eva K. Grebel$^{1}$
}
\address{%
Max-Planck-Institut f\"ur Astronomie, K\"onigstuhl 17, D-69117 Heidelberg,
      Germany
}

\begin{abstract}
I review the properties of the interstellar medium in the Magellanic Clouds
and Local Group dwarf galaxies.  The more massive, star-forming
galaxies show a complex, multi-phase ISM full of shells and holes
ranging from very cold phases (a few 10 K) to extremely hot
gas ($>10^6$ K).  
In environments with high UV radiation fields the formation of molecular
gas is suppressed, while in dwarfs with low UV fields molecular gas can
form at lower than typical Galactic column densities.
There is evidence of ongoing interactions, gas accretion, and 
stripping of gas in the Local Group dwarfs.  Some dwarf galaxies appear to be 
in various stages of transition from gas-rich to gas-poor systems.
No ISM was found in the least massive dwarfs.

\end{abstract}
\section{Introduction}
There are currently 36 known and probable
galaxy members of the Local Group within a zero-velocity surface of 1.2 Mpc.
Except for the three spiral galaxies M31, Milky Way, and M33, the 
irregular Large Magellanic Cloud, and the small elliptical galaxy M32
all other galaxies in the Local Group are dwarf galaxies with $M_V > -18$ mag.
In this review I will concentrate on the properties of the interstellar 
medium (ISM) of the Magellanic
Clouds and the Local Group dwarf galaxies, moving from gas-rich to gas-poor
objects.  Recent results from telescopes, instruments, and surveys such as 
{\em NANTEN}, {\em ATCA}, {\em HIPASS}, {\em JCMT},
{\em ISO}, {\em ROSAT}, {\em FUSE}, 
{\em ORFEUS}, {\em STIS}, Keck, and the {\em VLT} are significantly advancing 
our knowledge of the ISM in nearby dwarfs and provide an unprecedented,
multi-wavelength view of its different phases.
For a detailed review of the properties of all Local Group galaxies see
\cite{vdB00}. 

\section{The Magellanic Clouds}

\subsection{The Large Magellanic Cloud}

At high Galactic latitude and a distance of only $\sim$50 kpc, the
Large Magellanic Cloud (LMC) is one of the best-studied galaxies in the
Local Group.  With $0.5 \cdot 10^9$ M$_{\odot}$  
\cite{Kim98},
the gaseous component (neglecting He)
contributes $\sim 9$ \% to the total mass of the LMC
($5.3 \pm 1.0 \cdot 10^9$ M$_{\odot}$; \cite{Alves00}).  
The gas to
dust ratio is four times lower in the LMC than in the Milky Way 
\cite{Koorn82}.  
The total diffuse H$_2$ mass is $8\cdot 10^6$ M$_{\odot}$, $<2$\%
of the LMC's H\,{\sc i} mass and $\sim$1/9th of the Milky Way's
fraction \cite{Tumlin01}.
The reduced H$_2$ fraction may imply enhanced
destruction through UV photodissociation in low-metallicity environments
or suppressed H$_2$ on dust grains that H$_2$ 
\cite{Tumlin01}.  While high dust content is correlated with high H$_2$
abundances, H$_2$ does not trace CO or dust {\em per se}
\cite{Tumlin01}.

CO shows a strong correlation
with H\,{\sc ii} regions and young ($<10$ Myr) clusters,
but only little with 
older clusters and supernova remnants (SNRs) (\cite{Fukui99}, cf.\
\cite{Banas97}).  
Massive CO clouds have 
typical lifetimes of $\sim 6$ Myr and are dissipated within $\sim$3 Myr 
after the formation of young clusters.  CO clouds exist also
in quiescent areas without ongoing star formation; potential  
sites of future activity.  
Overall, the LMC clouds haver lower CO luminosities than in the
Milky Way and higher gas to dust ratios \cite{Fukui99}.
Individual cloud masses range from a few $10^4$ M$_{\odot}$
to $2\cdot 10^6$ M$_{\odot}$.  With 4 to $7\cdot 10^7$ M$_{\odot}$ the 
estimated molecular gas mass of the LMC amounts to 8 to 14\% 
of its total gas mass \cite{Fukui99}. 

H\,{\sc i} aperture synthesis maps of the LMC have revealed an ISM
with a turbulent, fractal structure that is self-similar on scales from 
tens to hundreds of pc \cite{Elme01}, 
likely due to the energy input of OB stars and supernova explosions.  The 
flocculent ISM consists of numerous shells and holes surrounded by broken
H\,{\sc i} filaments \cite{Kim98}.  At very 
large scales supershells dominate.  
23 H\,{\sc i} supershells (i.e., holes with sizes that exceed the 
H\,{\sc i} scale height) and 103 giant shells (sizes below the H\,{\sc i}
scale height) were identified \cite{Kim99}.  
Many of the giant shells interlock or collide with one 
another, or occur at the rims of supershells. They probably result from
winds of recently formed massive stars 
in a propagating-star-formation scenario.  Generally, the H\,{\sc i}
shells show little correlation with the optically dominant H\,{\sc ii}
shells, which suggests that H\,{\sc i} shells live longer than 
the OB stars that caused them initially \cite{Kim99}.  
H\,{\sc i} associated with H\,{\sc ii} typically exceeds the 
size of the ionized regions. 

The overall appearance of the H\,{\sc i}
disk of the LMC is symmetric, does not show obvious correlations with the 
optical bar, and reveals spiral features \cite{Kim98}.  Its southernmost
``spiral arm'' connects to the Magellanic Bridge, the tidal
H\,{\sc i} overdensity
between the LMC and the Small Magellanic Cloud (SMC).

Photoionization is the main contributor to the optical appearance of
the ISM at $\sim 10^4$ K
in the LMC and other gas-rich, star-forming galaxies.
The LMC has a total H$\alpha$ luminosity of $2.7\cdot10^{40}$ erg s$^{-1}$.
30 to 40\% are contributed by diffuse, extended gas 
\cite{Ken95}.  Nine H\,{\sc ii} supershells with diameters $>600$ 
pc are known in the LMC 
\cite{Meab80}.  Their rims are marked by strings of H\,{\sc ii} regions
and young clusters/OB associations.  The standard picture for supershells
suggests that these are expanding shells driven by propagating star
formation (e.g., \cite{McCray87}).  However, an age 
{\em gradient} consistent with this scenario was not
detected in the largest of these supershells, LMC4 
\cite{Dolphin98}.  Nor are the supershells
LMC1 \cite{Points00}, LMC2 \cite{Points99}, and LMC4 \cite{Dom95}
expanding as a whole, but instead appear to consist of hot gas confined 
between H\,{\sc i} sheets and show localized expansion.  
Supershells in several other galaxies
do not show evidence for expansion either \cite{Points99}, nor for
the expected young massive stellar populations \cite{Rhode99}.

More highly ionized gas can be effectively traced through ultraviolet
absorption lines from species such as 
such as C\,{\sc iv} ($10^5$ K), N\,{\sc v} ($2\cdot10^5$ K), and 
O\,{\sc vi} ($3\cdot10^5$ K; these temperatures are valid in the likely
case of collisionally ionized gas).  C\,{\sc iv} and O\,{\sc vi} 
are detected along sight lines across the entire LMC,
spatially uncorrelated with star-forming regions.  Its velocities indicate
that it is likely part of a hot, highly ionized corona of the LMC
(\cite{Wakker98}, \cite{Howk01}).  

Shock heating through fast stellar winds and, more importantly, supernova
explosions are the primary creation mechanisms for hot gas with $\ge 10^6$ K
\cite{Chu00}.  Diffuse hot gas in supershells, however, 
contributes only 6\% to the total X-ray emission from the LMC 
\cite{Points01}.  LMC2, the supershell east of 30\,Doradus,
has the highest X-ray surface brightness of all the supergiant shells in 
the LMC. The second highest X-ray surface
brightness comes from the yet unexplained extended ``spur'' south of LMC2.
The largest contribution to the LMC
X-ray budget comes from discrete X-ray binaries ($\sim 41$\%), followed
by diffuse field emission ($\sim 30$\%), and
discrete SNRs ($\sim 21$\%) \cite{Points01}.

Finally, the LMC is the only external galaxy detected thus far
in diffuse $\gamma$-rays, which are produced by (and directly proportional
to) the interaction of
cosmic rays (e.g., from supernovae) with the interstellar
medium \cite{Pav01}.  The integrated flux
above 100 MeV is $1.9 \cdot 10^{-7}$ photons cm$^{-2}$ s$^{-1}$ 
\cite{Sree92}.

\subsection{The Small Magellanic Cloud}

The SMC is the second most massive Milky Way
companion ($2\cdot10^9$ M$_{\odot}$; \cite{Wester97}).
With a total H\,{\sc i} mass of $4.2\cdot10^8$ M$_{\odot}$
\cite{Stan99}, 21\% of its mass 
are in the ISM.  The SMC's dust mass, on the other hand, is only $1.8\cdot10^4$
M$_{\odot}$ \cite{Stan00}, and its average dust to gas mass ratio is 
$8.2\cdot10^{-5}$, a factor 30 below the Galactic value.  The highest
concentrations of dust are found in luminous H\,{\sc ii} regions.
Cold gas appears to be mostly atomic rather than molecular due to the reduced
dust abundance, fewer coolants, and a higher UV radiation field (\cite{Stan00},
\cite{Dick00}), which help to photodissociate H$_2$.
Less than 15\% of the H\,{\sc i} is in cold gas, which tends to be colder
than in the Milky Way 
($\le 40$ K vs.\ 50 to 100 K \cite{Dick00}).
The diffuse H$_2$ mass is $2\cdot 10^6$ M$_{\odot}$, $\sim$0.5\% of
its H\,{\sc i} mass and 1/9th of the Galactic value, similar to the
reduced H$_2$ fraction in the LMC \cite{Tumlin01}.

Three H\,{\sc i} supershells ($>600$ pc) and 495 giant shells were 
detected in the SMC (\cite{Stave97}; 
\cite{Stan99}).  These shells appear to be
expanding.  Their rims coincide with a number of H\,{\sc ii} regions. 
Their centers lack pronounced H$\alpha$ emission in good 
agreement with their dynamical ages of $>10^7$ years and the propagating
star formation scenario proposed by \cite{McCray87}, though detailed
studies of the stellar age structure are lacking so far.  As in the LMC,
the ISM of the SMC is fractal \cite{Stan99}, likely due to turbulent energy
input.  The idea that the SMC consists of multiple components that are distinct
in location and velocity is not supported by the recent large-scale H\,{\sc i} 
data and was probably an artifact of the complex shell structure of the SMC
\cite{Stave97}.  On large scales, areas of high H\,{\sc i} column densities
coincide with the luminous  H\,{\sc ii} regions that form the bar and the 
wing of the SMC \cite{Stan99}.  The distribution of stars younger
than 200 Myr also traces these areas of recent massive star formation
well \cite{Zar00}.  Collisionally ionized gas with
a few $10^5$ K forms a hot halo around the SMC and shows
enhanced column densities toward star-forming regions 
\cite{Hoopes02}.  
Slightly enhanced diffuse X-ray emission has 
been detected along the SMC bar \cite{Snow99}.  

\subsection{The Magellanic Bridge and Stream}

The SMC has a distance of $\sim 60$ kpc from the Milky Way and $\sim 20$
kpc from the LMC.  SMC, LMC, and Milky Way interact tidally with each other,
which is reflected in, e.g., the H\,{\sc i} warp of the Milky Way disk 
\cite{Wein95}, in the thickening of the LMC's stellar  disk 
\cite{Wein00}, its elliptical extension toward the Milky Way 
\cite{vdM01}, in the star formation histories of the
three galaxies  (\cite{Gir95}; 
\cite{Greb99}; \cite{Rocha00}), and most notably
in the gaseous tidal features surrounding the Magellanic Clouds.  

The LMC and SMC are connected by the ``Magellanic Bridge'', an irregular, clumpy
H\,{\sc i} feature with a mass of $10^8$ M$_{\odot}$ that emanates
from both Clouds \cite{Putman98}. 
Cold (20 to 40 K) H\,{\sc i} gas has been detected in the Bridge
\cite{Kobul99}, and recent star formation 
occurred there over the past 10 to 25 Myr 
\cite{Demers98}.  Higher ionized species with temperatures up to $\sim 10^5$
K show an abundance pattern suggesting depletion into dust \cite{Lehner00}.  
Interestingly, stellar abundances in the Bridge 
were found to be $\sim -1.1$ dex \cite{Rolle99}, 
0.4 dex below the mean abundance of the young SMC population, which is
inconsistent with the proposed tidal origin 200 Myr ago \cite{Gard96}.  
However, it is conceivable that the Bridge formed from Magellanic Clouds 
material that mixed with an unenriched component \cite{Rolle99}, making
cloud-cloud collisions a possible star formation trigger \cite{Lehner00}.

Additional tidal H\,{\sc i} features include the leading arm ($10^7$ 
M$_{\odot}$, 25$^{\circ}$ length, \cite{Putman98}) and the patchy, clumpy
trailing arm ($10^{\circ} \times 100^{\circ}$) of the Magellanic Stream, 
in which no stars have been detected so far \cite{Putman00}.  The Magellanic
Stream is detected in H$\alpha$ due to photoionization by the Galaxy 
\cite{Bland99}.  The abundance patterns of interstellar absorption lines
are consistent with those in the SMC, and the H$_2$ detected in the leading
arm may originally have formed in the SMC \cite{Sembach01}.  Based on
their abundances, 
additional high-velocity clouds in the vicinity may have
been torn out of the SMC \cite{Lu98}.
 
\section{Other Local Group Irregulars}

The remaining Local Group dwarf irregulars (dIrrs) are more distant
from the dominant spirals, and fairly isolated.  While their star formation
activity and gas content generally decrease with decreasing galaxy mass,
their star formation histories and ISM properties present a less homogeneous
picture when considered in detail.  We first present the contrasting examples
of two comparatively high-mass dIrrs 
and then move on to the low mass end.

The H\,{\sc i} of IC\,10 (distance 660 kpc) is 7.2 times more extended than
its Holmberg radius \cite{Tomita93}.  While the inner part is a regularly
rotating disk full of shells and holes, the outer H\,{\sc i} gas is 
counterrotating \cite{Wilc98}.
IC\,10 is currently undergoing a massive starburst, which may be triggered 
and fueled by an infalling H\,{\sc i} cloud (\cite{Saito92}, \cite{Wilc98}).
Only upper
limits have been established for the diffuse X-ray emission \cite{RobWar00},
which may be due to the high foreground absorption toward IC\,10.
A non-thermal superbubble was detected that may be the 
result of several supernova explosions \cite{Yang93}.
The masses of the molecular clouds in IC\,10 are as high as 0.3 to $5\cdot10^6$
M$_\odot$ \cite{Petit98}.  Owing to the high radiation field and the 
destruction of small dust grains, the ratio of far-infrared 
[C\,{\sc ii}] to CO 1--0 emission is a
factor 4 larger than in the Milky Way \cite{Bolat00}, resulting in small CO
cores are surrounded by large [C\,{\sc ii}]-emitting envelopes \cite{Madden97}.
Two H$_2$O masers were detected in dense clouds in IC\,10, marking sites of
massive star formation \cite{Becker93}.  The internal dust content of IC\,10
is high \cite{Richer01}.

NGC\,6822, a dIrr at a distance of $\sim 500$ kpc, is also 
embedded in an  elongated H\,{\sc i} cloud with numerous shells and holes
that is much more extended than its stellar body \cite{deBlok00}.  Its
total  H\,{\sc i} mass is $1.1 \cdot 10^8$ M$_{\odot}$, $\sim 7$\% of its total
mass.  The masses of individual CO clouds reach up to 1 to $2\cdot10^5$ 
M$_{\odot}$ \cite{Petit98}, while the estimated H$_2$ content is 15\%
of the H\,{\sc i} mass \cite{Israel97}, and the dust to gas mass ratio is
$\sim 1.4 \cdot 10^{-4}$ \cite{Israel96}.  In comparison to IC\,10, NGC\,6822
is fairly quiescent, although it contains many H\,{\sc ii} regions.  Its
huge supershell ($2.0\times 1.4$ kpc) was likely caused by the passage of
and interaction with a nearby $10^7$  M$_{\odot}$ H\,{\sc i} cloud 
and does not show signs of expansion \cite{deBlok00}.

The H\,{\sc i} in low-mass dIrrs may be up to three times
more extended than the optical galaxy and is clumpy on
scales of 100 to 300 pc.
The most massive clumps reach $\sim 10^6$ 
M$_{\odot}$.  H\,{\sc i} concentrations tend to be close to
H\,{\sc ii} regions.  Some dIrrs contain cold H\,{\sc i} clouds
associated with molecular gas, while dIrrs without  
cold H\,{\sc i} also do not show ongoing star formation.
The total H\,{\sc i} masses are usually $< 10^8$
M$_{\odot}$.  The center of the H\,{\sc i} distribution coincides
roughly with the optical center of the dIrrs, although the H\,{\sc i} may
show a central depression surrounded by an H\,{\sc i} ring or arc (e.g.,
SagDIG, Leo\,A).
In contrast to the more massive dIrrs, the low-mass dIrrs show little 
to no rotation and appear to be dominated by chaotic motions. 
Details are given in \cite{Lo93}, \cite{YoungLo96}, \cite{YoungLo97a}, 
\cite{Elme00}. 

\section{Elliptical, Spheroidal, and Transition-Type Dwarfs}

The ISM of the dwarf elliptical (dE) companions of M31 exhibits puzzling
properties that are not yet understood.  In NGC 205, the stellar component
does not show rotation, while the H\,{\sc i} does \cite{YoungLo97b}.  In 
NGC\,185, stars and gas belong to the same kinematic system, while in NGC\,147
neither H\,{\sc i} nor molecular gas were 
detected \cite{YoungLo97b}.  In these three galaxies, the most recent 
measured star formation event took place 50 Myr, 100 Myr, and $>$1 Gyr ago,
respectively (\cite{Cappel99}, \cite{Mart99}, \cite{Han97}), but these
recent episodes cannot explain the differences in the ISM content.  With $10^6$
M$_{\odot}$, the amount of gas in NGC\,205 is a factor of 10 below what
one would expect from normal mass loss through stellar evolution, and the
kinematic differences between gas and stars make this an unlikely origin
\cite{Welch98}.  In NGC\,185 stellar mass loss may provide an explanation for
the H\,{\sc i}.
In both galaxies, the gas is less extended than the optical body, 
asymmetrically distributed, and clumpy on scales of less than 200 pc.
NGC\,205 contains CO and dust, which are closely associated with H\,{\sc i}
concentrations on scales of 100 pc \cite{Young00}.  While the molecular
clouds in NGC\,205 closely resemble Galactic clouds, the H\,{\sc i}
envelopes of the CO clouds have much lower column densities of only
$\sim 10^{20}$ cm$^{-2}$ due to the lower interstellar UV radiation field.
This implies that in dwarf galaxies with low UV radiation molecular gas
may form and survive even at column densities below $10^{21}$ cm$^{-2}$
as less shielding is required (\cite{Young00}; see also \cite{Elme00}).  
The gas to dust ratio, which
is similar to the one in the Milky Way, indicates that little dust gets
destroyed in this environment.  Even extremely cold dust with $<10$ K was
detected in NGC\,205 \cite{Haas98}. 

Owing to their prominent old stellar populations, some dIrrs resemble 
dwarf spheroidal (dSph) galaxies.  But while dSph galaxies appear to 
be devoid of gas, these galaxies have been detected in H\,{\sc i}.  
Thus they are classified as dIrr/dSph systems, galaxies that may be in
transition from low-mass dIrrs to gas-less dSphs.  LGS\,3, one of these
transition types, is at a distance of 280 kpc from M31 and experienced
low-rate star formation until 500 Myr ago \cite{Miller01}.  Its star
formation rate was not significant enough to expel its gas, and the 
H\,{\sc i} distribution is nicely centered on the optical galaxy.  In
contrast, the dIrr/dSph galaxy Phoenix (distance $\sim 400$ kpc from the
Milky Way), which formed stars continuously until 100 Myr ago \cite{Holtz00},  
does not contain H\,{\sc i} within the main body of the optical galaxy.
However, a nearby H\,{\sc i} cloud with $\sim 5 \cdot 10^6$ M$_{\odot}$ has 
a velocity consistent with its having originated from Phe (\cite{StG99},
\cite{Gallart01}).  
It may have been expelled through
supernova explosions (though its regular shape seems to argue against this),
or ram pressure stripping \cite{Gallart01}.  
Gas in two extended H\,{\sc i} lobes appears to be within
the tidal radius of Sculptor, a dSph without a young or
significant intermediate-age population \cite{Hurley99}, matching its
velocity  \cite{Car98}.  The amount of
gas detected is consistent with expectations from stellar mass loss 
through normal stellar evolution.  On the other hand, the surrounding field
is filled with similar clouds, suggesting the possibility of mere
coincidence \cite{Car99}. 
 
The upper limits for H\,{\sc i} in the other Local Group dSphs
are at column densities of a few $10^{17}$ cm$^{-2}$, well below
even of expectations from mass loss through normal stellar evolution.  At
earlier stages in their evolution, these dSphs were evidently capable of 
forming stars and retaining gas over extended periods of time. Even
those that are predominantly old show evidence for metallicity spreads
\cite{Grebel00}.  The Fornax dSph still formed stars $<200$ Myr ago 
\cite{GreSte99}, while sensitive H\,{\sc i} searches did not detect any gas 
\cite{Young99} -- perhaps a galaxy one step beyond Phe, just
having completed its transformation into a dSph?
Claims that gas exists at larger distances from dSphs
\cite{Blitz00} were not confirmed for several of them (e.g., Leo\,I 
\cite{Young00b}; And\,V \cite{Guha02}).  Searches for diffuse highly
ionized gas only yielded upper limits (Leo\,I: \cite{Bowen97}).
Internal effects such as supernova
explosions appear to be insufficient for removing the gas \cite{Mac99}.
Gas loss through tidal shocks
during perigalactic passages close to a massive galaxy may rid dSphs of
their gas \cite{Mayer01} and provide an explanation for transition-type
galaxies and the morphology-density relation, but environmental effects 
cannot explain isolated, gas-less dSphs like Tucana.   Hence many
interesting questions concerning the ISM in dwarf galaxies remain open.

\acknowledgements{It is a pleasure to thank the organizers for their kind
invitation, Y.-H.\ Chu and J.S.\ Gallagher for a critical
reading of the text, and Landessternwarte Heidelberg
for a quiet office where this review was completed. 
}

\vfill
\end{document}